\documentclass[preprint,showpacs,preprintnumbers,amsmath,amssymb,nofootinbib]{revtex4}

\usepackage{graphicx}
\usepackage{dcolumn}
\usepackage{bm}
\usepackage{epsfig}

\begin{document}


\title{Breathing mode in an improved transport approach}

\author{T. Gaitanos$^{1}$, A.B. Larionov$^{1,2}$, H. Lenske$^{1}$, U. Mosel$^{1}$}
\affiliation{$^{1}$Institut f\"{u}r Theoretische Physik,
Justus-Liebig-Universit\"{a}t Giessen,
D-35392 Giessen, Germany\\
$^{2}$Russian Research Center Kurchatov Institute,
RU-123182 Moscow, Russia}

\date{\today}

\begin{abstract}
The nuclear breathing-mode giant monopole resonance is studied within
an improved relativistic Boltzmann-Uehling-Uhlenbeck (BUU) transport
approach. As a new feature, the numerical treatment of ground state
nuclei and their phase-space evolution is realized with the same
semiclassical energy density functional.
With this new method a very good stability of ground state nuclei
in BUU simulations is achieved. This is important in extracting clear
breathing-mode signals for the excitation energy and, in particular,
for the lifetime from transport theoretical studies including mean-field
and collisional effects.
\end{abstract}



\maketitle


\section{\label{sec1}Introduction}

A giant monopole resonance (GMR), i.e. a collective isoscalar $0^+$
excitation of a nucleus, has been extensively investigated in
the past few decades, both theoretically and experimentally
(see Ref. \cite{HW01} for the most recent review).
It is nowadays well established, that the GMR is a nuclear
compression mode governed mostly by the incompressibility
modulus $K_\infty$ of nuclear matter \cite{blaizot}, which is
the key quantity for the description of nuclei, supernovae
explosions, neutron stars, and heavy-ion collisions.

Many microscopic models have been developed for the theoretical
description of giant resonances. They can be divided into two groups:
purely quantum mechanical approaches and semi-classical dynamical models.

The former group includes the models based on the nonrelativistic
\cite{Hamamoto,Sagawa} or relativistic \cite{Piekarewicz01,Piekarewicz09}
static Hartree-Fock plus random phase approximation (RPA) method.
The RPA technique can be derived from
a more general time-dependent Hartree-Fock (TDHF) theory
(c.f. \cite{Chomaz,vretenar1}) in the case of small-amplitude excitations.
The damping of a collective mode in a pure mean field RPA picture originates
from the coupling to the particle-hole excitations
(Landau damping or fragmentation width) and from coupling to the continuum
states, which is equivalent to the particle loss in TDHF calculations
\cite{Chomaz}.
Also constrained relativistic mean-field approaches have been developed
and applied in the case of a GMR \cite{vretenar1,stoitsov,Sharma09}.
The collective nature of the GMR in these quantum mechanical prescriptions
manifests itself from a coherent
superposition of many single-particle transitions from one major shell to
another \cite{fred}. Generally, within uncertainties of an underlying
energy-density functional, the fully quantum approaches are able to describe
the GMR energies for various sets of nuclei. However, the conclusions of
different authors on the GMR total width are clearly different.
Sagawa et al. \cite{Sagawa} claim that the Landau damping and coupling
to the continuum states explain the major part of the total GMR width
for the $^{208}$Pb nucleus and Sn isotopes. However,  Piekarewicz and Centelles
\cite{Piekarewicz09} state that the RPA calculation fails to account
for the spreading component of the full escape-plus-spreading width
because of the lack of coupling to the more complex than particle-hole
configurations.

The second group of models \cite{BBD89}-\cite{serbulent} solves the
BUU(-like) or Vlasov(-like) equations, which are the semiclassical limits of
a quantum kinetic equation \cite{BUU}.
An advantage of the kinetic transport equation with respect to the TDHF theory is
that binary collisions are naturally included, apart the interaction between
the particles due to the classical nuclear mean-field.
In the classical picture, a GMR can be qualitatively understood in terms of
a radial collective vibration of a nucleus: protons and neutrons oscillate
in phase with a certain amplitude, which is damped due to dissipation.
For this reason, the GMR is often referred to as a "breathing mode".
The frequency of the breathing mode characterizes the excitation energy and
the temporal damping of the amplitude --- the life time of the resonance.

Semi-classical treatments of a GMR in finite nuclei have been so far restricted
to pure Vlasov dynamics \cite{Abrosimov98,morawetz,serbulent}, where any collisional
effects are completely neglected. Thus, these models predicted the
excitation energy of the breathing mode appropriate well, but not the width.
Alternatively, the collisional effects in the semi-classical description of the
GMR were modeled within a linearized Landau-Vlasov equation \cite{KPS96},
which, however, takes into account the finite size effects only very approximately.
There are several works where the Boltzmann-Nordheim-Vlasov approach -- which
takes into account the selfconsistent nucleon mean field and nucleon-nucleon
collisions -- has been applied to the collective dipole \cite{Baran09,Pierr09} and
quadrupole \cite{LPCD00} motions excited in heavy-ion collisions at low energies.

The present work is an attempt to describe simultaneously the centroid
energy and the width of a GMR in finite nuclei. To this aim, we perform
the full BUU calculations taking into account both the mean field and
collision term. The results of the full BUU calculations are then compared
with the results of solution of the Vlasov equation to make the quantitative
conclusions on the contribution of two-body collisions to the total GMR width.

The numerical solution of a BUU equation for the case of a small-amplitude
collective vibration excited in a finite nuclear system is extremely difficult.
It requires a very good stability of the ground state configurations, which is
difficult to reach in a test-particle technique underlying any numerical
method to solve the BUU equation.  So far empirical density distributions have
been used to initialize ground state nuclei in transport theoretical simulations,
which might be not always consistent with the energy density functional used for
the propagation of the system. Another well known problem (see e.g. discussions
in Refs. \cite{BBD89,BDG90,SBD91}) is related to the calculation of Pauli
blocking factors in the Uehling-Uhlenbeck collision integral for the small-amplitude
Fermi surface distortions.

We thus have improved the relativistic transport approach \cite{LBGM07,LMSG08,Gaitanos08}
based on the Giessen Boltzmann-Uehling-Uhlenbeck (GiBUU) transport model \cite{GiBUU}
by performing relativistic Thomas-Fermi (RTF) calculations with the same
energy-density functional as that used in the dynamical evolution.
In this context, the isovector-vector $\rho$ meson field together with its gradient
terms have been included in the calculations of the present work.
The initialization of neutron and proton densities according to the
RTF calculation largely improves the ground-state stability
in numerical simulations of the Vlasov and BUU dynamics. We have also worked out
to improve the numerical treatment of the Pauli blocking in test-particle simulations
of the small-amplitude nuclear motions, in-particular, in the nuclear surface regions.

The structure of the work is as follows: the standard theoretical background is
presented in section \ref{sec2}. The modified initialization procedure
and numerical treatment of the Pauli blocking are then presented in section
\ref{sec3}. In section \ref{sec4},
results of the ground state simulations are discussed, before the
calculations of the GMR excitations are presented. Finally, conclusions and outlook are
presented in the last section \ref{sec5}.

\section{\label{sec2}The relativistic transport equation}

The baryonic mean-field is modeled within the non-linear Walecka model
(mean-field approximation of the QHD) \cite{QHD97,Boguta77}.
The non-linear Walecka model Lagrangian includes the nucleon
field $\psi$, the isoscalar-scalar $\sigma$ meson field, isoscalar-vector
$\omega$ meson field, isovector-vector $\vec{\rho}$ meson field and the
electromagnetic field ${\cal A}$ and reads ($\hbar=c=1$) as:
\begin{eqnarray}
{\cal L}
& = &
\bar\psi [
\gamma_{\mu}
( i\partial^{\mu} - g_{\omega}\omega^{\mu} - g_{\rho}\vec{\tau}\vec{\rho}^{\,\mu}
- e\frac{1+\tau_3}{2} {\cal A}^{\mu})
- ( m + g_{\sigma} \sigma) ] \psi
\nonumber \\
& + &
{1 \over 2} \partial_{\mu}\sigma \partial^{\mu}\sigma - U(\sigma)
-  {1 \over 4} \Omega_{\mu\nu} \Omega^{\mu\nu}
+ {1 \over 2} m_\omega^2 \omega^2
\nonumber \\
& - &  {1 \over 4} \vec{R}_{\mu\nu} \vec{R}^{\mu\nu}
+ {1 \over 2} m_\rho^2 \vec{\rho}^{\,2}
-  {1 \over 4} F_{\mu\nu} F^{\mu\nu}~,
\label{Lagr}
\end{eqnarray}
where $\Omega_{\mu\nu} = \partial_\mu \omega_\nu - \partial_\nu \omega_\mu$,
$\vec{R}_{\mu\nu} = \partial_\mu \vec{\rho}_\nu - \partial_\nu \vec{\rho}_\mu$
and $F_{\mu\nu} = \partial_\mu {\cal A}_\nu - \partial_\nu {\cal A}_\mu$ are the
field tensors. In Eq. (\ref{Lagr}), the arrow above a symbol indicates the isovector
character of the corresponding field. The term $U(\sigma)={1 \over 2} m_\sigma^2 \sigma^2
+{1 \over 3} g_2 \sigma^3 +{1 \over 4} g_3 \sigma^4$ contains the
selfinteractions of the $\sigma$ field, added according to \cite{Boguta77}.
The bare hadron masses $m,~m_\sigma,~m_\omega$ and $m_\rho$, coupling constants
$g_\sigma,~g_\omega,~g_\rho$ and the non-linear parameters $g_{2}$ and $g_{3}$
have been adopted from the $NL3^*$ parametrization of the non-linear Walecka model,
which gives reasonable values for the incompressibility
modulus, $K_\infty=258$ MeV, and the nucleon Dirac effective mass
$m^*=0.594m$ at the saturation density $\rho_0=0.150$ fm$^{-3}$ of nuclear matter
\cite{LalaNew}. The $NL3^*$ parametrization is the modification of
the well known $NL3$ set of parameters \cite{LaLa} adjusted to describe
the ground state properties of both spherical and deformed nuclei as
well as the GMR energies of heavy nuclei. The momentum dependence of the
proton-nucleus optical potential, studied recently in Ref. \cite{NLD}, is
of minor importance here.

The theoretical description of heavy-ion collisions is realized
within the GiBUU transport
approach \cite{GiBUU}, which is based on a relativistic kinetic
equation. Thorough derivations of the transport equation from an
effective hadron-meson field theory \cite{QHD97}, can be found
elsewhere \cite{blaettel}. The relativistic kinetic
equation reads
\begin{eqnarray}
& & (p^{*0})^{-1}\left[
p^{*\mu} \partial_{\mu}^{x} + \left( p^{*}_{\mu} F_i^{k\mu}
+ m^{*} \partial_{x}^{k} m^{*}  \right)
\partial_{k}^{p^{*}}
\right] f_{i}(x,{\bf p}^{*}) = \sum_{j=n,p} I_{ij}
\label{rbuu}
\end{eqnarray}
with $\mu=0,1,2,3$ and $k=1,2,3$. The l.h.s. of Eq. (\ref{rbuu}) describes
the classical Vlasov propagation of a one-body phase space distribution function
$f_{i}(x,{\bf p}^{*})$ for protons and neutrons ($i=p,n$) in the mean meson
fields.  This is expressed in terms of a kinetic four-momentum $p^{*}=p_i-V_i$,
where $p_i$ is the canonical four-momentum, of the field tensor
$F_i^{\mu\nu} = \partial^\mu V_i^\nu - \partial^\nu V_i^\mu$, and of the
Dirac effective mass $m^* = m + S$. Here $V_i^\mu$ and $S$ are the
vector\footnote{Assuming that there is no mixing between proton
and neutron states, the 1-st and 2-nd isospin components of the $\rho$-meson fields
vanish.} and scalar field, respectively.
\begin{eqnarray}
   V_i^\mu & = & g_\omega \omega^\mu + g_\rho \tau^3_i \rho^{3\mu}
         + \frac{e}{2} (1+\tau^3_i) {\cal A}^\mu~,
\label{V^mu}\\
   S & = & g_\sigma \sigma~,
\label{S}
\end{eqnarray}
where $\tau^3_i=+(-)1$ for $i=p(n)$.

The particles are assumed to be on the Dirac effective mass shell, i.e.
\begin{equation}
   p^{*0}=\sqrt{m^{*2}+{\bf p}^{*2}}~.   \label{mass_shell}
\end{equation}

In fact, the GiBUU model propagates not only nucleons, but also all resonances
up to the mass of $2$ GeV, as well as mesons, e.g., pions, kaons. However, in
the present study we will concentrate only on the nucleonic degrees of freedom.
In this case, the collision terms in the r.h.s. of Eq. (\ref{rbuu}) are
the Uehling-Uhlenbeck collision integrals describing elastic
nucleon-nucleon scattering:
\begin{eqnarray}
   I_{ij} &=&\int\, \frac{ 2 d^3 p_2^\star }{ (2\pi)^3 }
                                           \int\,
d\sigma_{ij}({\bf p}^*_1,{\bf p}^*_2;{\bf p}^*_{1^\prime},{\bf p}^*_{2^\prime}) %
                                           \, v_{12}  \nonumber \\
 &\times&  [\, f_i({\bf p}^*_{1^\prime})\, f_j({\bf p}^*_{2^\prime})\,
             \bar{f}_i({\bf p}^*_1) \, \bar{f}_j({\bf p}^*_2)
    - f_i({\bf p}^*_1) \, f_j({\bf p}^*_2) \,
      \bar{f}_i({\bf p}^*_{1^\prime}) \, \bar{f}_j({\bf p}^*_{2^\prime})\,]~,
\label{I_ij}
\end{eqnarray}
where the time-space argument $x$ of the distribution functions is dropped for
brevity and ${\bf p}^*_1 \equiv {\bf p}^*$. The hole distribution functions
are denoted as $\bar{f}_i({\bf p}^*)\equiv(1-f_i({\bf p}^*))$.
Thus the final state Pauli blocking is explicitly
included in (\ref{I_ij}). The collision integral (\ref{I_ij})
depends on the differential elastic scattering cross section
$d\sigma_{ij}({\bf p}^*_1,{\bf p}^*_2;{\bf p}^*_{1^\prime},{\bf p}^*_{2^\prime})$
with initial momenta ${\bf p}^*_1,{\bf p}^*_2$ and final momenta
${\bf p}^*_{1^\prime},{\bf p}^*_{2^\prime}$,
and on the relative velocity $v_{12}$ of colliding nucleons.
We use in this work the energy- and angular-dependent vacuum nucleon-nucleon
cross sections (see \cite{LBGM07} for details).

An exact solution of the set of the coupled transport equations
for the different hadrons is not possible. Therefore, the commonly used
test-particle method for the Vlasov part is applied, whereas the
collision integral is modeled in a parallel-ensemble Monte-Carlo algorithm.
The GiBUU transport model has not been applied to low energy reactions so far.
An important requirement here is a very good description of the
ground state of a nucleus in test particle method, which is the topic of
the next chapter.

\section{\label{sec3}Improved method for nuclear ground states in GiBUU}


Practically all existing nuclear kinetic transport models based on the
test particle technique (c.f. \cite{morawetz,serbulent,Baran09,GiBUU} and refs. therein),
including GiBUU, use the same method of the nuclear ground state preparation.
Typically, the coordinates of test particles are sampled according to empirical
Woods-Saxon or even uniform density profiles, while the test particle momenta are
distributed with the help of a local density approximation.
The standard numerical treatment of the transport equation works well for
high energy reactions, in which several collective flow observables are described
quantitatively well \cite{FOPI}. However, in low energy reactions the memory
of the exit channel to the initial configuration is important. A
disadvantage of standard numerical treatments is that
the initial distribution functions of protons and neutrons deviate from
the corresponding static solutions of the Vlasov equation.
Therefore, nuclei are not initialized in their proper ground states.
This makes their dynamical propagation unstable. Another
source of spurious instability is the numerical treatment of the Pauli blocking,
which mainly influences the momentum space of the test particles in full BUU
simulations.

\subsection{\label{sec3a} Phase space initialization}

We improve the phase space initialization of ground state nuclei in the transport
model in the following way: the nuclear ground state is
described in a semi-classical treatment, in which the density distribution of
a spherical ground state nuclei is obtained by minimizing the energy functional.
In RMF, the energy functional corresponds to the relativistic Hamiltonian density,
which is obtained from the $T^{00}$-component of the energy-momentum tensor.
The same functional is then used for the propagation of the system.
In calculations of the time evolution, we neglect the time derivatives of the mean
mesonic fields and of the electromagnetic field in the Lagrangian (\ref{Lagr}).
For the mean mesonic fields, this should be a reasonable approximation, since
the time scale of the GMR motion $\sim40-80$ fm/c (c.f. Fig.~\ref{fig5} below)
is much larger than the time scale of the free field oscillations
$\sim 1/m_{\rm mes}=0.2-0.4$ fm/c, where $m_{\rm mes}\sim0.5-0.8$ GeV is
a meson mass.
For the electromagnetic field, neglecting time derivatives
corresponds to disregarding a radiation. The last could be, in-principle,
treated in terms of a quantum transition probability. However, in the specific
case of $0^+$ transitions, the electromagnetic radiation is suppressed \cite{QED}.
The space components of electromagnetic field are also
neglected, i.e. ${\cal A}\equiv({\cal A}^0,0,0,0)$. In the most calculations,
in order to save CPU time, we also drop the space components of the vector
meson fields. However, for completeness, the formalism below takes into
account these components.

The RMF Hamiltonian density of the non-linear Walecka model reads:
\begin{eqnarray}
\epsilon \equiv T^{00} & = &
\frac{2}{(2\pi)^{3}}\sum_{i=p,n} \int d^3p^* p_i^0(x,{\bf p}^*) f_i(x,{\bf p}^*)
+ \frac{1}{2}(\nabla\sigma)^2 + U(\sigma)
\nonumber\\
& - & \frac{1}{2}
[ \nabla\omega^\mu\nabla\omega_\mu + m_{\omega}^{2}\omega^2
 +\nabla\rho^{3\mu}\nabla\rho^3_{~\mu} + m_{\rho}^{2}(\rho^3)^2
 +  (\nabla  {\cal A}^0)^2 ]~.
\label{T00}
\end{eqnarray}

The Hamiltonian density (\ref{T00}) takes into account the gradients of
the meson fields in coordinate space according to \cite{LMSG08}.
This is important for description of surface effects, which
is impossible in the usual local density approximation
neglecting the gradient terms (c.f. \cite{serbulent,LBGM07}).
In distinction to the previous GiBUU calculations in the RMF mode,
the isovector $\rho$-meson field and the Coulomb field ${\cal A}_{0}$ are
explicitly included now.

The meson and electromagnetic field equations have the following
form
\begin{eqnarray}
     & & \left( -\triangle + { \partial U(\sigma) \over \partial \sigma } \right)\,\sigma
      = - g_\sigma \rho_S~,                   \label{eqSigma} \\
     & & (-\triangle + m_\omega^2)\,\omega^\mu
      =   g_\omega j_B^\mu~,                  \label{eqOmega} \\
     & & (-\triangle + m_\rho^2)\,\rho^{3\mu}
      =   g_\rho j_I^\mu~,                    \label{eqRho} \\
     & & -\triangle {\cal A}^\mu
      =  e j_p^\mu~.                          \label{eqEM}
\end{eqnarray}
The source densities and currents in the r.h.s. of Eqs. (\ref{eqSigma})-(\ref{eqEM})
are expressed in terms of the distribution functions as follows
\begin{eqnarray}
\rho_{S}(x)  & = & \frac{2}{(2\pi)^3} \sum_{i=p,n} \int d^3p^*
\frac{m^*}{p^{*0}} f_i(x,{\bf p}^*)~,                  \label{rho_s} \\
j_i^\mu(x) & = & \frac{2}{(2\pi)^3} \int d^3p^* \frac{ p^{*\mu} }{ p^{*0} }
                 f_i(x,{\bf p}^*)~,~~~~(i=p,n)         \label{j_i} \\
j_B^\mu(x) & = &   j_p^\mu(x) + j_n^\mu(x)~,           \label{j_B} \\
j_I^\mu(x) & = &   j_p^\mu(x) - j_n^\mu(x)~.           \label{j_I}
\end{eqnarray}

In the static case, the distribution functions are Fermi distributions:
\begin{equation}
   f_i^{\rm static}({\bf r},{\bf p}^*) = \Theta(p_{F_i}({\bf r})-|{\bf p}^*|)~.
                                            \label{staticDF}
\end{equation}
The space components of the source currents (\ref{j_i})-(\ref{j_I})
as well as those of the mesonic and electromagnetic fields disappear
in a static system. Therefore, the Hamiltonian density (\ref{T00})
becomes a functional of proton and neutron densities
$\rho_{p,n}=j_{p,n}^{0}$ only.

The RTF equations for a static nucleus
with $Z$ protons, $N$ neutrons and $A=N+Z$ nucleons are obtained
by applying a variational principle to the total energy
$E=\int \epsilon[\rho_p,\rho_n] d^3r$ under the constraint of particle
number conservation:
\begin{equation}
\delta
       \int
	\left(
		\epsilon[\rho_p,\rho_n]
               -\mu_{p}\rho_{p}({\bf r})
               -\mu_{n}\rho_{n}({\bf r})
	\right) d^3r
   = 0~.
\label{var}
\end{equation}
The chemical potentials for protons and neutrons, $\mu_{p,n}$,
are fixed by the conditions
\begin{equation}
  Z = \int \rho_p({\bf r}) d^3r~,~~~
  N = \int \rho_n({\bf r}) d^3r~.
\label{chempot}
\end{equation}
Substituting the Hamiltonian density functional (\ref{T00}) in
Eq. (\ref{var}) leads to the RTF-equations for protons
and neutrons:
\begin{eqnarray}
   g_{\omega}\omega^0 + g_{\rho}\rho^{30}
                    + e{\cal A}^0 + E_{F_p}^{*} & = & \mu_{p}~,  \label{spEnergies_p} \\
   g_{\omega}\omega^0 - g_{\rho}\rho^{30}
                    + E_{F_n}^{*} & = & \mu_{n} ~,
\label{spEnergies_n}
\end{eqnarray}
where $E_{F_{p,n}}^{*}=\sqrt{p_{F_{p,n}}^{2}+m^{*2}}$.
For a spherical nucleus, the RTF equations (\ref{spEnergies_p},\ref{spEnergies_n}) together
with the field equations (\ref{eqSigma})-(\ref{eqEM}) completely
determine the radial dependence of the proton and neutron densities
and fields, i.e. $\rho_{p,n}(r)$, $\sigma(r)$, $\omega^0(r)$,
$\rho^{30}(r)$ and ${\cal A}^0(r)$. The densities are obtained as the
selfconsistent solution of
Eqs. (\ref{eqSigma}-\ref{eqEM},\ref{chempot}-\ref{spEnergies_n}),
with $p_{F_{i}}=\left( 3\pi^{2}\rho_{i} \right)^{1/3}$ ($i=p,n$) according to the
local density approximation. This method then yields a consistent description of the
groundstate. The densities are realistic but somewhat too steep in the surface region.

The solution of the RTF equations gives the nuclear density
$\rho_{i}(r)$ and
the local Fermi-momentum $p_{F_{i}}(r)$ ($i=p,n$). The test particles for
the numerical solution of the transport equation are distributed according to these
functions. The propagation of the system is described by the
test particle equations of motion, which directly follow
from the transport equation (\ref{rbuu}) by putting its r.h.s. equal
to zero and read ($j=1,...,A \cdot {\cal N}$ with ${\cal N}$ being
the number of test particles per nucleon):
\begin{eqnarray}
&& \dot{\bf r}_j =  \frac{{\bf p}_j^*}{p_j^{*0}}        \label{rDot} \\
&& \dot{p}^{*k}_j =
\frac{p^{*}_{j\mu}}{p^{*0}_j}F_j^{k\mu}
+ \frac{m_j^*}{p^{*0}_j} \partial^k m^{*}_{j}         \label{pDot}~.
\end{eqnarray}
At each time step of the simulation, the scalar density (\ref{rho_s})
and currents (\ref{j_i})-(\ref{j_I}) are calculated on a grid in coordinate
space.
Using these quantities, the equations of motion for the meson and Coulomb
fields (\ref{eqSigma})-(\ref{eqEM}) are solved numerically. Note, that
the solution of Eq. (\ref{eqSigma}) requires some more iterations, since the scalar
density $\rho_{S}$ in its r.h.s depends itself on the $\sigma$ meson field via the
effective mass $m^{*}$.

The collision term in the r.h.s. of Eq.(\ref{rbuu}) is simulated by
explicit two-body collisions between test particles using the geometrical
collision criterium (c.f. \cite{LBGM07} for details). An important
feature here is the numerical treatment of the Pauli blocking, which is
discussed in detail  below.

\subsection{\label{sec3b} Pauli blocking}

The frequency of two-body collisions in a Fermi gas depends on the
occupancies of the scattering final states via the hole distribution
functions $\bar f_i$ (see Eq.(\ref{I_ij})). By the energy and momentum
conservation, no collisions take place at zero temperature, when
$f_i({\bf r},{\bf p}^\star)=\theta(p_{F_i}({\bf r})-|{\bf p}^\star|)$.
In test-particle simulations, however, it is impossible to model the
exact $T=0$ Fermi distribution. This causes some spurious two-body
collisions even in the ground state nucleus. The magnitude of this
spurious effect crucially depends on a numerical technique of
the Pauli blocking calculation. In the standard GiBUU
\cite{GiBUU,BussPhD}, the occupation number $f_i({\bf r},{\bf p}^\star)$
is calculated by counting the number of test particles in the phase
space volume element composed of small spherical volumes
$\Delta V_r$ with radius $r_r$ centered at ${\bf r}$ in coordinate space
and $\Delta V_p$ with radius $r_p$ centered at ${\bf p}^\star$ in
momentum space:
\begin{equation}
   f_i({\bf r},{\bf p}^\star) = \sum_{j:~{\bf p}_j^\star \in \Delta V_p}
   \frac{1}{\kappa(2\pi\sigma^2)^{3/2}}
   \int\limits_{\Delta V_r, |{\bf r}-{\bf r}_j| < r_c}\,d^3r
   \exp\left\{ -\frac{({\bf r}-{\bf r}_j)^2}{2\sigma^2}\right\}~, \label{f_i}
\end{equation}
where
\begin{equation}
   \kappa = \frac{2\, \Delta V_r\, \Delta V_p\, {\cal N}}{(2\pi)^3}
   \frac{4\pi}{(2\pi\sigma^2)^{3/2}}
   \int\limits_0^{r_c} dr\, r^2 \exp\left\{-\frac{r^2}{2\sigma^2}\right\}
                                                                 \label{kappa}
\end{equation}
is a normalization factor. In Eq.(\ref{f_i}), the sum is taken over all test particles
$j$ of the type $i=p,n$ whose momenta belong to the volume $\Delta V_p$. In the coordinate
space, the test particles are represented by Gaussians of the width $\sigma$ cutted-off
at the radial distance $r_c$. The default values of parameters are
$r_p=80$ MeV/c, $r_r=1.86$ fm, $\sigma=1$ fm, $r_c=2.2$ fm. This set of parameters is
a compromise between the quality of the Pauli blocking in the
ground state and the smallness of statistical fluctuations in the case of simulations
with ${\cal N}\sim200$ test particles per nucleon. Typically, this is good enough for modeling
the heavy-ion collisions at the beam energies above $\sim100$ MeV/nucleon. In the case of a
small-amplitude dynamics near nuclear ground state, the accuracy provided by
Eqs.(\ref{f_i}),(\ref{kappa}) is not enough, when the default parameters are used.
The main reason is the constant, i.e. momentum-independent radius $r_p$, which
introduces a spurious temperature of the order of several MeV.
To reduce this effect, we have introduced the coordinate- and momentum-dependent
radius of the momentum space volume $\Delta V_p$ as
$r_p({\bf r},|{\bf p}^\star|)=\mbox{max}(20~\mbox{MeV/c}, p_{F_i}({\bf r}) - |{\bf p}^\star|)$,
which provides a sharper momentum dependence near Fermi momentum. The calculations
of the present work are performed with ${\cal N}=1000-10000$. This allows us to use the
reduced parameters also in the coordinate space: $r_r=0.9-1.86$ fm, $\sigma=0.5$ fm,
$r_c=1.1$ fm. The calculations with the default and modified set of parameters
for the Pauli blocking are compared in the following Sect.~\ref{sec4}.

\section{\label{sec4}Results}


\begin{figure}[t]
\begin{center}
\includegraphics[clip=true,width=0.75\columnwidth,angle=0.]
{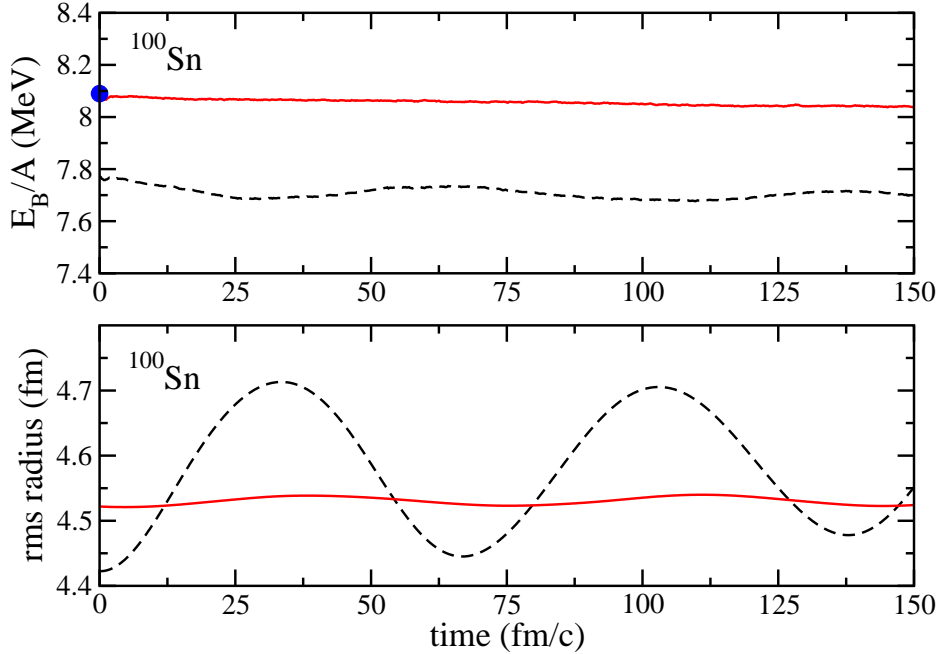}
\caption{\label{fig1} 
(Color online) Time evolution of the binding energy per nucleon
(panel on the top) and root mean square (rms) radius (panel on the bottom) for a
ground state ${}^{100}$Sn nucleus. Vlasov calculations using the
(dashed) standard initialization and the
(solid) improved initialization are shown. The filled circle in the top panel
at $t=0~fm/c$ gives the RTF-value of the binding energy.
}
\end{center}
\end{figure}


We study first the influence of the initialization
method on the temporal evolution of nuclei in their ground states.
The improved transport model is applied then to the dynamics of low energy
nuclear excitations, which are simulated by initializing slightly
expanded nuclei. We present results from pure Vlasov and full BUU
calculations for different nuclei. If not indicated elsewhere,
for the Vlasov calculations $10000$ test particles per nucleon were used.
The full BUU calculations were performed with $1000$ test particles due to
time limitations.

\subsection{\label{sec4a} Stability of the ground state}

Fig. \ref{fig1} shows the time evolution of the binding energy per
nucleon and the root mean square (rms) radius of a ground state
$^{100}$Sn nucleus. The standard initialization method using
the empirical Woods-Saxon density distribution produces
the binding energy smaller by 0.3-0.4 MeV/nucleon with respect
to the RTF value of $E_{B}/A \simeq 8.1$ MeV. This is expected,
since the minimum of the total energy is not reached by the
standard initialization. The binding energy varies with time
due to numerical errors in the solution of the time evolution
equations (\ref{rDot}),(\ref{pDot}) and field equations
(\ref{eqSigma})-(\ref{eqEM}). For the standard initialization,
the rms radius reveals quite strong fluctuations, comparable in
the amplitude with the true GMR vibrations (see Fig.~\ref{fig4}).
These artificial temporal oscillations lead also to a significant
particle loss with increasing time, if collisions are included (see below).
Applying the improved initialization, in which the same
Hamiltonian density functional is used for both the initialization of
the nucleus and its temporal propagation, the situation becomes considerably
better. At $t=0$ fm/c the value of the binding energy per nucleon
agrees with the corresponding RTF value, and the rms radius stays
almost constant in time.


\begin{figure}[t]
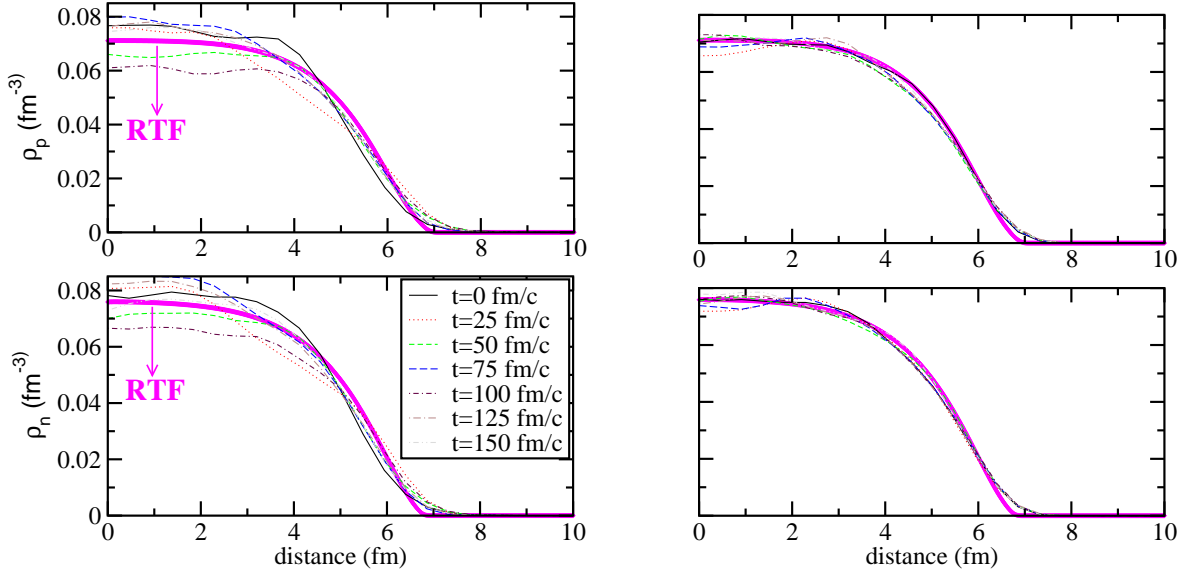

\begin{center}
\includegraphics[clip=true,width=0.47\columnwidth,angle=0.]
{Fig2a.eps}
\includegraphics[clip=true,width=0.47\columnwidth,angle=0.]
{Fig2b.eps}
\caption{\label{fig2} 
(Color online) Density profiles of protons (panels on the top) and neutrons
(panels on the bottom) for the same nucleus
as in Fig. \protect\ref{fig1}. The thick curves are RTF calculations.
The other curves show density distributions from the Vlasov dynamics
at different times (as indicated) using the standard initialization
(panels on the left) and the improved one (panels on the right).
}
\end{center}
\end{figure}

A more detailed picture of the Vlasov calculations with the standard
and the improved initialization method is shown in Fig. \ref{fig2}
in terms of the proton and neutron density distributions. Using the
standard initialization (figures on the left panel) the initial
($t=0~fm/c$) density profiles do not fit that one of RTF.
This leads to significant density oscillations around the true
ground state density profiles (RTF) with the result of spurious
oscillations in the rms-radius of the system (see again
Fig. \ref{fig1}). A consistent treatment between the ground state
nucleus and its propagation leads to very good stable configurations,
see graphs on the right panel of Fig. \ref{fig2}.


\begin{figure}[t]
\begin{center}
\includegraphics[clip=true,width=0.75\columnwidth,angle=0.]
{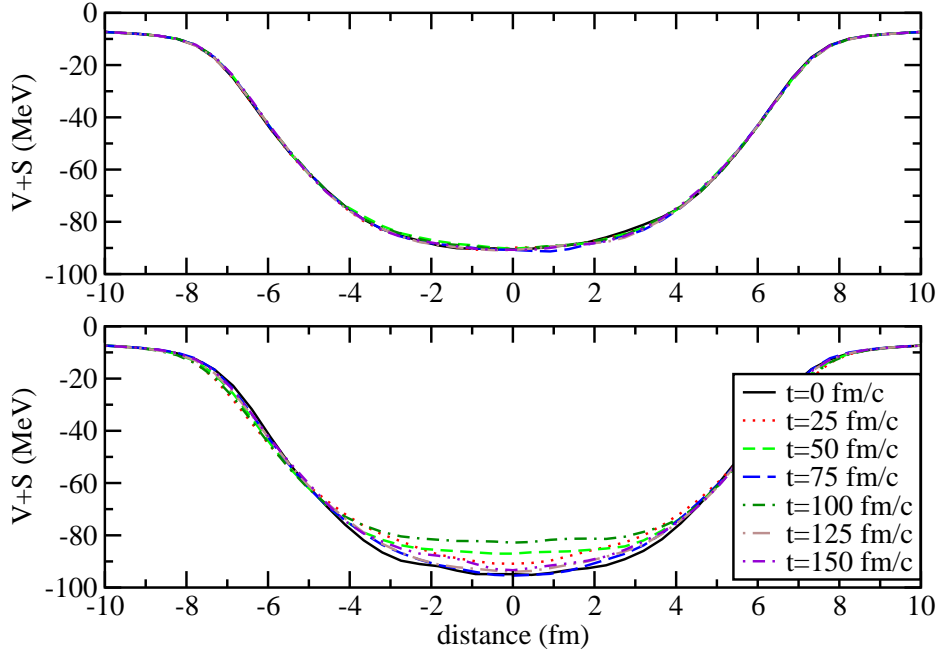}
\caption{\label{fig3} 
(Color online) The proton mean field potential
$V^0+S$  (see Eqs. (\protect\ref{V^mu},~\protect\ref{S}))
along the $z$-axis passing through the center of the
$^{100}$Sn nucleus. The different curves show the
Vlasov results at different times (as indicated)
using the improved initialization (upper panel) and the
standard one (lower panel).
}
\end{center}
\end{figure}

We remind that in relativistic transport studies the central mean-field
potential arises from the sum of the large negative Lorentz scalar
and large positive Lorentz vector potentials. Thus, small spurious
variations in density cause strong numerical fluctuations in the mean-field
potential.
This is demonstrated in Fig. \ref{fig3}, where the
mean field potential is displayed as a function of the coordinate along
the central $z$-axis.
The Vlasov calculations with the standard initialization (panel on the
bottom) show large fluctuations of the order of $10\%$, while these
fluctuations almost vanish in the calculations using the improved
initialization method.


\begin{figure}[t]
\begin{center}
\includegraphics[clip=true,width=0.7\columnwidth,angle=0.]
{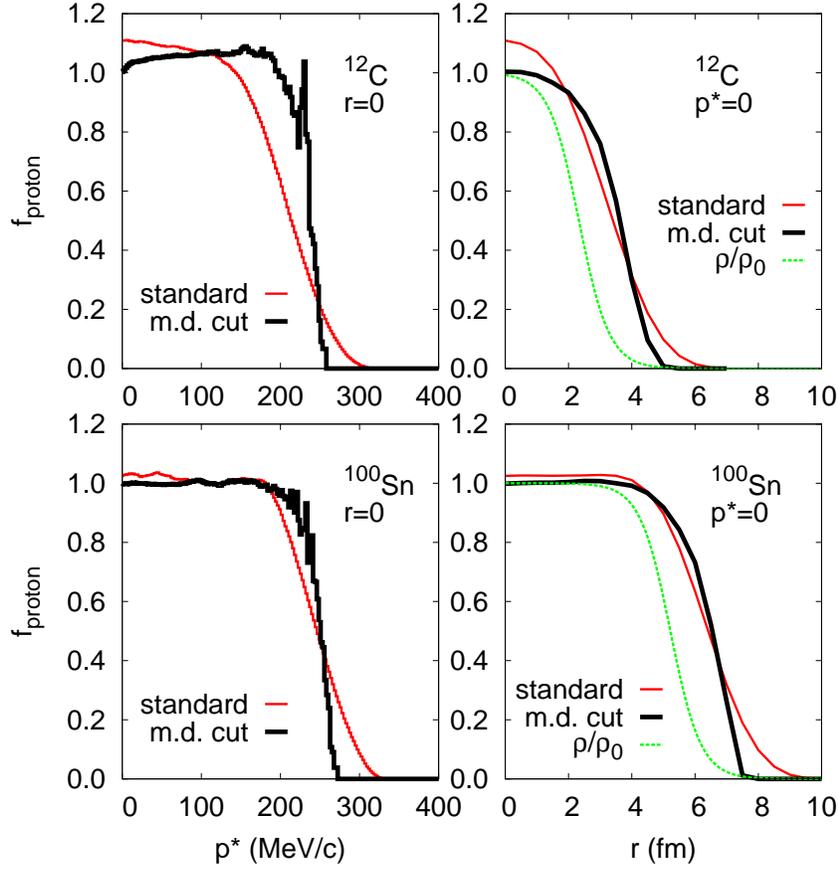}
\caption{\label{fig:pauli} 
(Color online) Momentum dependence of the proton
occupation number $f_{\rm proton}$ in the center of $^{12}$C and $^{100}$Sn
nuclei is shown in the upper and lower left panels, respectively.
Radial dependence of $f_{\rm proton}$ at the zero momentum
for $^{12}$C and $^{100}$Sn is depicted in the upper and lower right panels,
respectively. The results are presented for the standard (thin solid lines)
and momentum-dependent (thick solid lines) radius $r_p$. The fluctuations
of $f_{\rm proton}$ near Fermi momentum $p_F\simeq250$ MeV/c are due to
finite number of test particles per nucleon which was set to 10000 in this
calculation. The nucleon density in units of $\rho_0$ is shown additionally
by dashed lines in the right panels. We see that at the half-central-density
radius, the proton occupation number is only about 10\% below unity.
}
\end{center}
\end{figure}


Including two-body collisions requires a careful implementation of the
Pauli blocking to prevent the ground state to be destroyed. To give
an impression of how well it is working in our test particle calculations,
Fig.~\ref{fig:pauli} shows the momentum (left) and radial (right)
dependence of the proton occupation numbers, which are used
in the evaluation of the Pauli blocking factors, for $^{12}$C and
$^{100}$Sn nuclei. The calculation with the default parameters of
a Pauli blocking produces a rather diffuse momentum dependence,
especially for the light $^{12}$C nucleus. Using the momentum-dependent
radius $r_p$ and the reduced width of a Gaussian, as explained in
subsect. \ref{sec3b}, largely improves the momentum dependence
of occupation numbers near Fermi momentum. The radial dependence
of the occupation numbers also becomes closer to the
step function, when the calculation is done with the modified
Pauli blocking parameters.


\begin{figure}[t]
\begin{center}
\includegraphics[clip=true,width=0.7\columnwidth,angle=0.]
{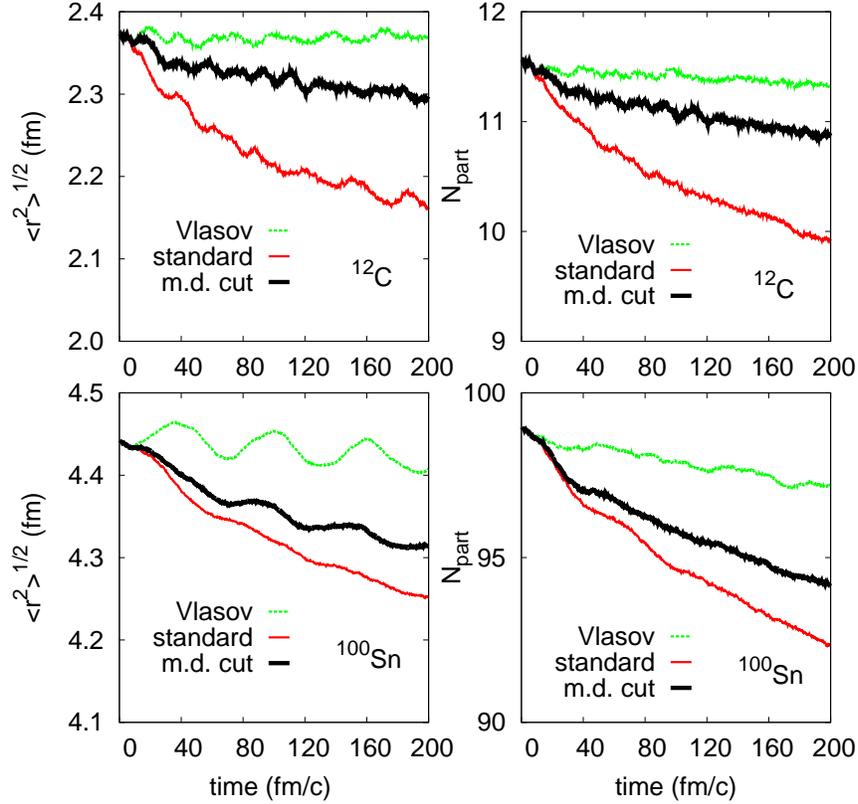}
\caption{\label{fig:r_rms} 
(Color online) The nuclear rms-radius Eq.(\ref{rms})
(left panels) and the number of particles in the high-density
region Eq.(\ref{N_part}) (right panels) as a function of time
for the ground state $^{12}$C and $^{100}$Sn nuclei.
The Vlasov calculations are represented by dashed lines.
The full BUU results with the standard and modified Pauli
blocking parameters are shown by thin and thick solid
lines, respectively.
}
\end{center}
\end{figure}


To demonstrate the effect of Pauli blocking parameters on the ground
state evolution, in Fig.~\ref{fig:r_rms}, we present the time dependence
of the rms-radius of a nucleus (left) and of the number of particles in the
high-density, $\rho >  \rho_{\rm min}$, space region (right) for the carbon
and tin nuclei. Explicitly, these quantities have been calculated as
\begin{eqnarray}
   N_{\rm part} &=& \int\limits_{\rho > \rho_{\rm min}} d^3r \,
                    \rho ({\bf r}) \quad ,  \label{N_part}\\
   <r^{2}>  &=& N_{\rm part}^{-1}\int\limits_{\rho > \rho_{\rm min}} d^3r \,
                    r^{2} \, \rho({\bf r}) \label{rms}
\end{eqnarray}
with $\rho_{\rm min}=0.1\rho_0$. Vlasov calculations practically conserve
the number of particles in the high density region and produce almost
constant in time rms-radii. The spurious effect of two-body collisions
in the ground state nuclei leads to the particle emission to vacuum
which amounts after 200 fm/c to about 10\% of the total mass number
in the case of standard Pauli blocking parameters and $\sim 5$\% in the
case of the modified parameters. Correspondingly, the rms-radius gets
reduced. This spurious reduction has to be excluded in the calculation
of the rms-radius in full BUU simulations, as discussed later on.
Below, we will always apply the modified Pauli blocking parameters
as explained in subsect. \ref{sec3b}.

\subsection{\label{sec4b} Giant Monopole Resonance: GiBUU calculations}

The different methods of initialization of nuclear ground states influence
the dynamical calculations of excited nuclei. We will model the
low energy nuclear giant monopole collective excitations by initializing an
expanded nucleus at $t=0~fm/c$. This is realized by re-scaling
the coordinates of the test particles such that the corresponding
excitation energy is close to the experimental values. The relation between
the scaling parameter and the excitation energy is obtained by expanding the
energy per nucleon $E/A$ around saturation density or the ground state
radius $R_{0}$:
\begin{eqnarray}
E/A(R) \backsimeq E_{0}/A + \frac{1}{2} K_\infty
\left(
    \frac{R-R_0}{R_0}
\right)^{2}
\label{scaling}
\quad .
\end{eqnarray}
With $K_\infty=258~MeV$ and using the experimental values \cite{blaizot} for the
excitation energy $\Delta E=E-E_{0}$ one obtains scaling parameters
$\Delta R/R = (R-R_{0})/R_{0}$ in the range of $\approx 0.03-0.05$ for $A \in (56,208)$.
The system is then propagated either without (Vlasov mode) or
with (BUU mode) collisions between the nucleons of the excited nucleus.

\begin{figure}[t]
\begin{center}
\includegraphics[clip=true,width=0.7\columnwidth,angle=0.]
{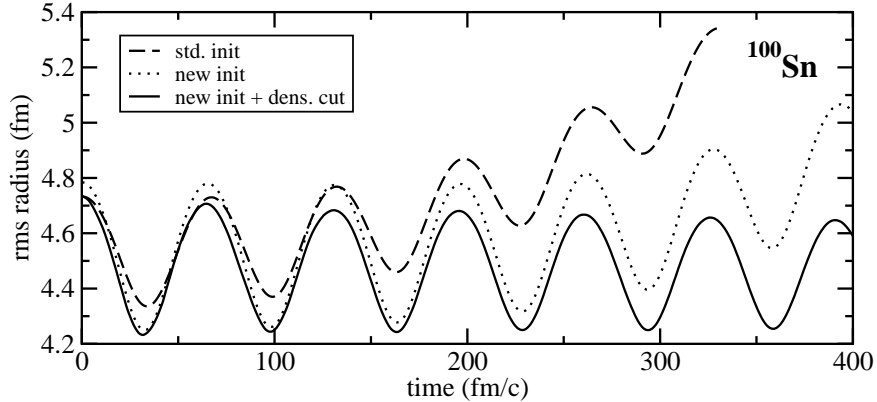}
\caption{\label{fig4} 
(Color online) Time dependence of the rms-radius for the
excited ${}^{100}$Sn nucleus. The meaning of the different curves is
explained in the text. The calculations are done in the Vlasov mode.
}
\end{center}
\end{figure}

Fig. \ref{fig4} shows the time evolution of a ${}^{100}$Sn nucleus
using the different prescriptions of initialization.

The standard method does not provide stable solutions and the rms-radius of
the $^{100}$Sn-nucleus explodes for $t>200~fm/c$ due to strong particle loss.
The situation is considerably improved in the new initialization. A clear
oscillation signal can be seen with a particle emission for very
late times ($t>275~fm/c$) which is, however, only moderate.
Emitted particles increase the rms-radius of the total nuclear system
which hinders the true oscillation signal.
They have to be excluded, therefore, in the calculation of the rms-radius
of an oscillating nucleus when extracting the excitation energy and
width of the GMR. We thus consider only particles at densities
higher than $0.1\rho_0$ according to Eqs.(\ref{N_part}),(\ref{rms}).
After this correction we obtain a clear signal for the resonance. The
period of the oscillation characterizes the frequency and thus the
excitation energy of the resonance, and an exponential damping (see below)
--- its finite lifetime.


\begin{figure}[t]
\begin{center}
\includegraphics[clip=true,width=0.7\columnwidth,angle=0.]
{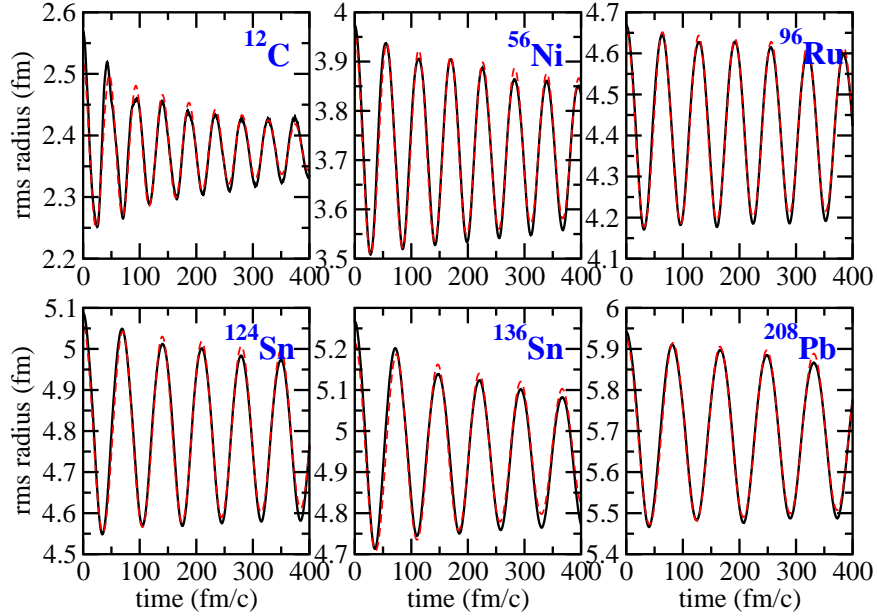}
\caption{\label{fig5} 
(Color online) Time dependence of the rms-radius for different
nuclei, as indicated. (Solid) Vlasov calculations, (dashed)
fit according to Eq. (\ref{fit}).
}
\end{center}
\end{figure}
We have performed pure Vlasov mode calculations for different excited
nuclei and analyzed the results in terms of the time dependence of the
rms-radius, as seen in Fig. \ref{fig5}. With increasing mass number
the frequency of the oscillation decreases and thus also the excitation
energy. An exponential damping is visible, even without
the inclusion of collisional effects. This effect has been interpreted
as a wall friction \cite{Blocki78,Sierk78}, and it will be discussed later.
An important feature in the study of nuclear collective excitations is
the calculation of the lifetime or the width of the different multipole
modes of the nuclear excitation. TDHF theory and Vlasov dynamics
does not include any collisional broadening effects.

The nuclear collective dynamics of giant multipole vibrations in
the ground state nuclei has not been so far investigated
within the full BUU equation, mainly due to the reasons of a ground state
instability. With the improved initialization method the Vlasov propagation of a ground
state is almost perfect. The numerical procedure of the Pauli blocking in
a full BUU ground state simulation is improved by applying the modifying method,
but is still far from being exact.
This situation leads to a spurious particle emission in ground state BUU
calculations and thus to a spurious escape width $\Gamma_{es}$, which
impedes the determination of the total width.

\begin{figure}[t]
\begin{center}
\includegraphics[clip=true,width=0.7\columnwidth,angle=0.]
{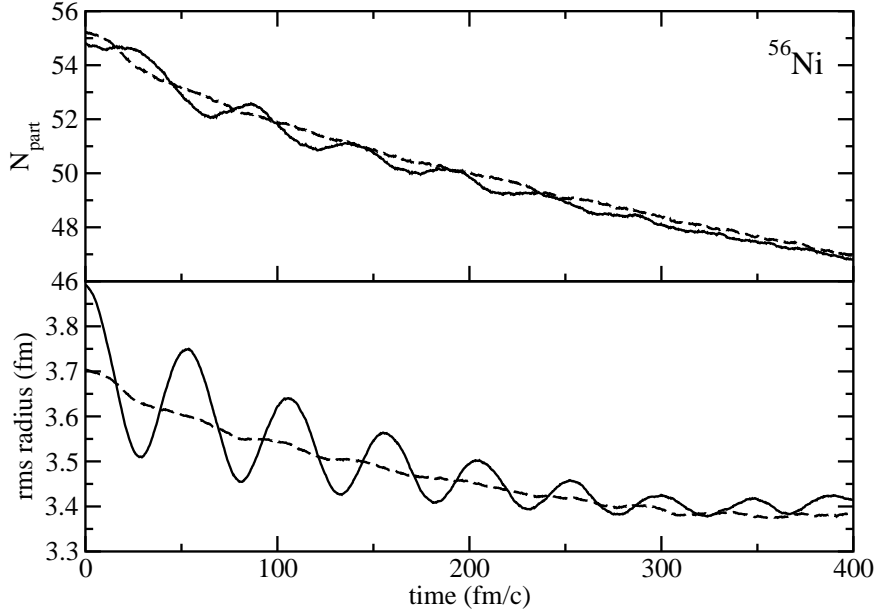}
\caption{\label{fig5a} 
(Color online) Time dependence of the number of particles in
high-density region (top panel) and the r.m.s. radius (bottom panel)
defined according to Eqs. (\ref{N_part}),(\ref{rms}) for a ${}^{56}$Ni nucleus.
Dashed lines: ground state BUU calculation, solid lines: BUU calculation
for an excited ${}^{56}$Ni nucleus in the GMR mode.
}
\end{center}
\end{figure}
The spurious contribution to the total width has therefore to be excluded, as
explained in Fig. \ref{fig5a}, where the time evolution of
bound particles for a nucleus in its ground state (dashed curve)
and in the GMR mode (solid curve) is shown. Already in the ground state, particles
leave the nucleus due to spurious two-body collisions resulting in a
spurious escape width $\Gamma_{es}$. On the other hand, the BUU simulation for the
excited nucleus contains also an escape width which is very similar to that of the
ground state. Asumming $\Gamma_{es}$ to be the same for both ground state and
excited state we can make the following Ansatz:
\begin{eqnarray}
N_{gs}(t) & = & N_{0}\exp(-\Gamma_{es} t)
\quad\mbox{for the ground state}\quad
\nonumber\\
N_{exc}(t) & = & N_{0}\exp(-\Gamma_{es} t)F(t)
\quad\mbox{for the excited GMR state}\quad
\quad ,
\label{esc}
\end{eqnarray}
where $N_{0}$ is the number of particles at $t=0$ fm/c and $F(t)$ is a fit function
for the oscillation signal, which contains the stochastic collisional
width:
\begin{equation}
F(t) = \alpha + \beta \cos( \omega t +\delta ) \exp(-\gamma t)
\quad .
\label{fit}
\end{equation}
The spurious width can be excluded by taking the ratio of $N_{gs}$ and $N_{exc}$,
or their difference in the case of very small value of $\Gamma_{es}$. The latter method
is applied here in extracting the GMR width, since $\Gamma_{es}\sim 0.044,0.045$ MeV for
a ground state and an excited nucleus, respectively.
Thus, the physical escape width (relative to that in the ground state) is
almost negligible, and we obtain the total width in full BUU calculations by a
fit according Eq. (\ref{fit}) to the so-called corrected r.m.s. radius
defined as
\begin{equation}
 <r^2_{corr}>^{1/2}(t) = <r^{2}_{exc}>^{1/2}(t) - <r^{2}_{gs}>^{1/2}(t)
                       + <r^{2}_{gs}>^{1/2}(0)~,  \label{r^2_corr}
\end{equation}
where the depletion due to spurious particle loss is subtracted.
However, the damping rate $\gamma$ of the corrected r.m.s. radius oscillations
is still somewhat influenced by an imperfect Pauli blocking procedure, which
makes some uncertainty in our calculations of the GMR width. 
We also have checked the consistency of the extracted results on $E^{*}$ 
and $\Gamma$ obtained with the subtraction method by performing additional 
Fourier analyses of $<r^2 >^{1/2}(t)$ before and after the correction. 
Lorentzian fits to the Fourier spectra lead to values for the width, which 
are the same with those 
extracted with the subtraction method. Also the excitation energy is not 
essentially affected. However, the excitation energy in full BUU calculations 
is slightly above the corresponding values in Vlasov mode. This effect is due 
to the spurious particle emission in full BUU calculations, which is furthermore 
related to the collective response of a smaller system.


\begin{figure}[t]
\begin{center}
\includegraphics[clip=true,width=0.7\columnwidth,angle=0.]
{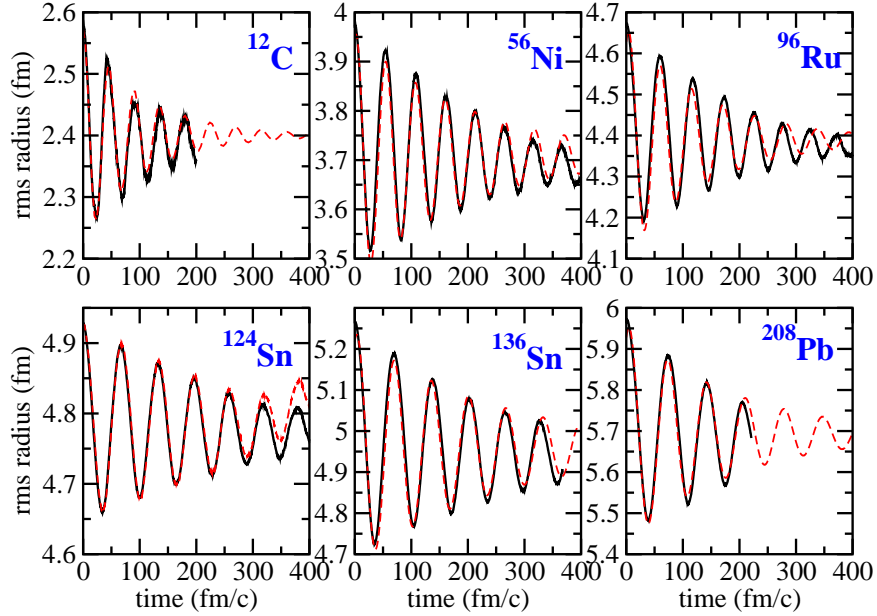}
\caption{\label{fig6} 
(Color online) The corrected r.m.s. radius of Eq. (\ref{r^2_corr})
plotted as a function of time for different nuclei --- solid lines.
Calculations are done in the full BUU mode.
The fit according to Eq. (\ref{fit}) is shown by the dashed lines.
}
\end{center}
\end{figure}
The results for the  corrected r.m.s. radius produced by full BUU calculations
for different nuclei are shown in Fig. \ref{fig6}. Again, a clear oscillation
signal is visible, however, due to the inclusion of collisions, a significant
damping of the oscillations of the rms-radii appears, as compared to the pure
Vlasov calculations of Fig.~\ref{fig5}. We note, that in the case of a pure Vlasov mode
calculation, the difference between corrected and not corrected r.m.s. radii
is negligibly small.

We have performed a fit to the results displayed in Figs. \ref{fig5} and \ref{fig6}
according to Eq. (\ref{fit}) to obtain the centroid energies and the total width of
the breathing mode. The results are summarized in Fig. \ref{fig7} in terms
of the excitation energy $E^\star=\omega$ and the damping width
$\Gamma=2\gamma$ of the GMR, as a function of the mass number. The
factor of two in the last formula is motivated by our intention to report
the Full Width at Half Maximum (FWHM) of the GMR strength \cite{KPS96}.
The gray band in the transport calculations for the width is an estimation
related to the numerical uncertainty of the Pauli blocking factors.
We first discuss the results obtained for $E^\star$ and then those for the
width.

\subsection{\label{sec4c} Giant Monopole Resonance: discussion}

The theoretical Vlasov-calculations for the excitation energy scale with
$A^{-1/3}$ and overestimate moderately the experimental data. The
mass dependence is also consistent with TDHF calculations and with
microscopic RPA studies \cite{blaizot}. Similar results for the excitation
energy are obtained using full BUU calculations. It is a well known question,
why relativistic structure calculations can explain the excitation energies of
GMR with a higher value of the incompressibility modulus $K_\infty=250-270$ MeV
(c.f. refs. \cite{LalaNew,Sharma09,Piekarewicz02,Vretenar03}) than
$K_\infty\simeq220$ MeV deduced from nonrelativistic approaches \cite{blaizot}
It has been shown in Ref. \cite{Piekarewicz02}, that this is at least partly related
to a stiffer density behaviour of the symmetry energy in RMF models with respect to
the Skyrme-type effective interactions. On the other hand, the calculations of Ref.
\cite{Sharma09} have revealed the influence of differences in the surface compressibility
in different RMF models on the GMR frequency. There is also another possible reason
for differences between RMF and nonrelativistic approaches, which we address below.

Our calculations have been performed without nuclear Lorentz force,
i.e., without taking into account the space-like components of the vector field.

\begin{table}[t]
\begin{center}
\caption{\label{table1} \small{
Excitation energy $E^{\star}$ for a ${}^{208}$Pb nucleus in different models:
(Vlasov wLF) Vlasov calculation including the Lorentz force, (Vlasov)
Vlasov calculation without the Lorentz force, (RPA) RPA calculations by
Lalazissis et al. \protect\cite{LalaNew}. All the results are given in the case
of NL3$^{*}$ model ($K_\infty=258$ MeV).
}}
\vskip 0.5cm
\begin{tabular}{|l|c|c|c|c|c|}
\hline\hline
 model  & Vlasov & Vlasov wLF & RPA & exp. \\
\hline\hline
$E^{\star}$ (MeV) & 15.2 & 14.2 & 13.9 & 13.7$\pm$ 0.5 \\
\hline\hline
\end{tabular}
\end{center}
\vskip 0.5cm
\vskip 0.5cm
\end{table}
We expect that a nonrelativistic Vlasov calculation, employing the same energy-density
functional suitably parameterized, e.g. in a Skyrme form,  will produce very
similar results. To study the influence of the nuclear Lorentz force on the excitation
energy (and the width) of the GMR, a Vlasov calculation explicitly taking into account
the space-like components of the vector field has been performed for the lead-208 nucleus.
The results are shown in table \ref{table1} and compared with other theoretical
models of nuclear structure. First of all, the effect of the Lorentz force is
negligible for the GMR width (not shown in the table). The excitation energy is
moderately affected. In particular, a decrease of $E^{\star}$ by $\sim 6.5 \%$
towards the experimental data and the relativistic structure calculations is
observed, when the Lorentz force is included in the Vlasov calculation. We note
that the same value for the compression modulus has been used in both Vlasov
calculations. This result indicates the importance of the genuine relativistic
effects when extracting the incompressibility modulus from GMR studies.

\begin{figure}[t]
\begin{center}
\includegraphics[clip=true,width=0.7\columnwidth,angle=0.]
{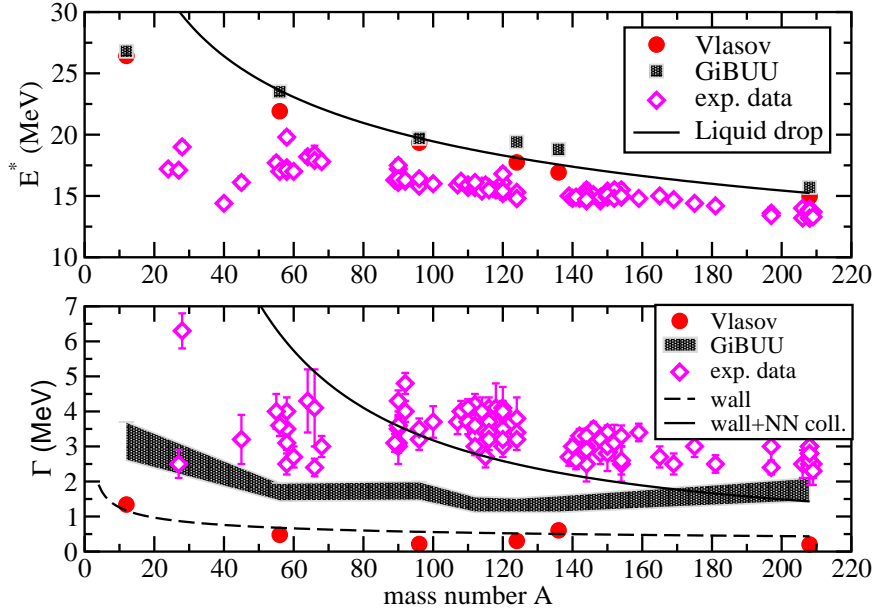}
\caption{\label{fig7} 
(Color online) Excitation energy (upper panel)
and width (lower panel) of the GMR as a function of the mass number.
Vlasov and full GiBUU results (as indicated)
are compared with experimental data (open diamonds). The solid line in the upper
panel shows the fit by Eq.(\ref{E^star}) with $\eta=90$.
In the lower panel, the dashed line depicts the fit of the Vlasov calculation using
the wall formula (\ref{tau_wall}) with $\xi=20$, while the solid line shows the full
width $\Gamma=-2\omega_I$ (see Eq.(\ref{omega_I})) taking into account both the wall
and collisional contributions. Experimental data are from Ref. \cite{Youngblood}.
}
\end{center}
\end{figure}

Another feature of interest in the transport results using both modes, full BUU and 
Vlasov, is the moderate decrease of the monopole frequencies with increasing neutron 
excess, as one can see in the results for $E^{*}$ of Fig.~\ref{fig7} (upper panel) for 
the Sn-isotopes. Such a trend is also supported experimentally \cite{LGL07}. It is 
well known that the monopole frequencies are affected by the slope of the symmetry 
energy around saturation, as in detailed discussed in Ref. \cite{baran}. It would 
be a challenge to extend this transport study to particularly isospin asymmetric 
systems, such as the Sn-isotopes, where systematic experimental studies 
exist \cite{LGL07}, and to investigate more exotic collective modes in neutron-rich 
systems.

To understand the GiBUU results on the mass dependence of the GMR
centroid energy on a qualitative level, we have performed some simple estimations. 
As it is well known from empirical GMR systematics 
\cite{HW01}, the centroid energy of the GMR follows the $A^{-1/3}$ law:
\begin{equation}
E^\star = \eta A^{-1/3}     \label{E^star} \quad .
\end{equation}
This behaviour can be understood as a consequence of a 
sound-like excitation in a finite system, i.e. $E^\star = v_s k$,
where $v_s$ is a sound velocity and $k=\pi/R$ is the eigenvalue 
of the lowest compressional mode determined from the disappearance 
of the pressure at the free surface \cite{BM}, $R=1.2A^{1/3}$
is the nuclear radius. The hydrodynamical model \cite{BM} would give 
$v_s=(K_\infty/9m)^{1/2}\simeq0.17$, which is the first sound velocity. 
Here, we have used the value of the incompressibility modulus 
$K_\infty=258$ MeV provided by the NL3$^*$ model. The dependence 
$\eta \propto K_A^{1/2}$, where $K_A$ is, however, the incompressibility modulus 
of the {\it finite nucleus}, is also provided by the scaling model of the 
GMR \cite{blaizot}. This model is usually applied in extraction of $K_A$
from experimental data on GMR energy (c.f. \cite{LGL07}) by using
the relation $E^\star = \sqrt{K_A/m<r^2>}$, where $<r^2> \simeq 3R^2/5$ is the r.m.s.
radius of the nucleus.

On the other hand,
according to the Fermi liquid theory \cite{AK59} the low-temperature
collective excitations in the infinite system are of the zero sound 
type. It is well known, that the propagating zero sound type solutions 
of the dispersion relation for collisionless Fermi liquid at zero temperature 
exist only for the repulsive particle-hole interactions (c.f. \cite{blaizot}),
which is not the case for the NL3$^*$ interaction used in the present work
(see below).
Introduction of finite temperature, generally, restore the collectivity for 
attractive particle-hole interactions, although it was proved only for 
momentum-independent interactions \cite{KLD97}. It is quite difficult, 
therefore, to actually identify the GMR vibration as the zero sound mode.
Assuming, nevertheless, the nuclear matter zero sound nature of the 
GMR vibration, we can estimate the sound velocity as $v_s \simeq v_F =p_F/m_L^\star \simeq 0.42$, 
where $p_F=257$ MeV/c is the Fermi momentum at the nuclear matter saturation density 
and $m_L^* = \sqrt{ m^{*2} + p_F^2}=0.65m$ is the Landau effective mass in the case
of the NL3$^*$ model.  

As a result, one gets the values 90, 111 and 223 
for the coefficient $\eta$ in Eq. (\ref{E^star}) for the hydrodynamical,
scaling and zero sound pictures of the GMR, respectively. In the case of
the scaling model, we assumed that $K_A=K_\infty$, which is a quite
rough assumption. Generally, one has $K_A < K_\infty$ mostly
due to the surface contribution \cite{Sharma09}.  
It turns out (see the upper panel of Fig.~\ref{fig7}), that
Eq. (\ref{E^star}) with the ``hydrodynamical'' value of $\eta=90$   
well fits the Vlasov results for large masses numbers
$A \geq 100$, although the reason for this is not fully clear for us.

For light nuclei, as one can see from Fig.~\ref{fig7}, the 
transport calculations  overestimate the experimental data on $E^\star$. 
This is the region where the RTF method becomes unreliable because the 
surface properties are not well described. However, the experimental determination
of the GMR parameters in light nuclei is rather uncertain due to strong 
fragmentation of the $0^{+}$ strength \cite{HW01}.


The situation for the width is more involved. Pure Vlasov calculations predict
a very small value for the GMR width and do not fit the data, as expected.
The inclusion of collisions in the full BUU calculations improves the
comparison between theory and experiment considerably. The
underprediction of the theoretical calculations to the data
becomes smaller, but an exact agreement is not achieved.
For a deeper interpretation of the GiBUU results on the mass dependence of the
GMR width several additional analytical calculations were performed.


The width of a collective vibration is related to the dissipation processes.
Within a pure mean field Vlasov calculation, the only damping mechanism is
the one-body dissipation governed by a coupling of the single-particle and
collective motions. Specifically for finite systems, damping arises due to
collisions of particles with a moving wall. This one-body dissipation
leads to the following ``wall'' formula for the collective energy
dissipation rate \cite{Blocki78}
\begin{equation}
   \dot E = m \rho\,\overline{v} \oint \dot n^2 d\sigma
\label{wall_formula}
\quad ,
\end{equation}
where $\rho=\rho_{p}+\rho_{n}$ is the nucleon density, $\dot n$ is the normal
component of the wall velocity, and the integral
is taken over the surface of a vessel. $\overline{v}$ is
the average speed of particles in the vessel.
For cold nuclear matter one has $\overline{v}= \frac{3}{4} v_F$.
The formula (\ref{wall_formula}) is derived under strong simplifying assumptions
of the gas at rest inside the vessel and of a sharp potential wall driven through
the vessel. These conditions are usually assumed to be valid for small surface
vibrations of the incompressible nuclear droplet \cite{Blocki78}, possibly
with some modifications \cite{Sierk78} (see also \cite{KPS96} and refs. therein).
Application of the wall formula to the compressional modes is more
questionable. Nevertheless by performing a rather simple calculation for the
GMR mode within the liquid-drop model with a free surface we arrived to the
usual formula for the one-body relaxation time (c.f. \cite{Blocki78,KPS96})
\begin{equation}
   \tau_{\rm wall} = \frac{2R}{\overline{v}} \xi
\label{tau_wall}
\quad ,
\end{equation}
with $\xi=0.5$ which means an extremely strong dissipation. E.g.,
for $^{208}$Pb we obtain the wall dissipation contribution to the GMR
width $\Gamma_{\rm wall}=2/\tau_{\rm wall} \simeq 17$ MeV.
This result has to be considered as an extremely rough approximation. The
self-consistency corrections, i.e. the collective motion of nucleons
near the surface region \cite{Sierk78} will modify Eq. (\ref{wall_formula}) and
$\tau_{\rm wall}$. Other effects like the surface diffuseness and curvature
are not taken into account by the wall formula at all. Having these reservations
in mind we will treat $\xi$ in Eq. (\ref{tau_wall}) as a free parameter and
determine it from comparison with the results of calculations in a Vlasov mode.
This produces the value $\xi\simeq20$ as demonstrated in the
lower panel of Fig.~\ref{fig7}.

We will now discuss the two-body dissipation.
It follows from the Uehling-Uhlenbeck collision integral that
the two-body collisions take place only in the case of local deviations
of the Fermi surface from a spherical shape or/and in the case of a finite
temperature. Moreover, the relaxation rate of a nonequilibrated
Fermi gas toward thermal equilibrium depends mainly on the total
excitation energy and not on the concrete shape of a Fermi
surface deformation \cite{Bertsch78}.
In particular, in the linear approximation with respect to the
deviation of a distribution function from the local equilibrium the
relaxation rate is proportional to $T^2$ \cite{AK59,BS70}.
This means, that collisional damping of a small Fermi
surface distortions practically vanishes at $T=0$.

Thus collisional damping of a vibrational motion excited in
the ground state nuclear system is practically absent at the beginning
of time evolution and gradually switches on as some part of the collective
vibrational energy is transferred to the heat. In Refs. \cite{BDG90,SBD91},
an ``apparent temperature'' has been introduced in the calculation of
the collisional widths of the giant quadrupole and giant dipole vibrations
built on the ground state nuclei.
The ``apparent temperature'' has been extracted in \cite{BDG90,SBD91} by
subtracting the collective energy $E_{\rm coll}$ of a vibrational mode
from the total excitation energy $E^\star$:
$T=\sqrt{(E^\star-E_{\rm coll})/a}$, where $a$ is a level density parameter.
This ``temperature'' has the real
physical meaning only when the system reached a complete thermal
equilibrium. Since the Pauli blocking factors in the
Uehling-Uhlenbeck collision integral depend crucially on a temperature,
calculations taking into account the time-dependent ``apparent temperature''
strongly increase the spreading width of a collective mode
\cite{BDG90,SBD91} to a good  agreement with experiment.
In our estimates of the two-body dissipation we will use the upper
limit for the ``temperature'' by putting $E_{\rm coll}=0$.
For the level density parameter we will use the Fermi gas expression
$a=\pi^2 A/4E_F$.

The GMR mode can be considered as a sound-like excitation inducing
the Fermi surface distortions of all multipolarities
$l \geq 2$ \cite{KPS96}. At $T \ll E_F$
the relaxation time of a small amplitude Fermi surface distortion
with multipolarity $l \geq 2$ is (c.f. \cite{AK59,BS70,KPS96})
$\tau_l = \kappa_l/T^2$ where $\kappa_l$ depends on the NN
cross sections. The parameters
$\kappa_l$ have been computed in \cite{DKL99,LCBD99} both for isovector and
isoscalar vibrations for various choices of the NN cross sections.
In the case of realistic energy and angular dependent
vacuum NN cross sections this resulted in the following values
of the isoscalar parameters: $\kappa_l=868,~881$ and $401$ MeV$^2$fm/c
for $l=2,~3$ and $+\infty$, respectively. The case $l=+\infty$ corresponds
to the relaxation of a single particle-hole configuration.
For simplicity, we will set the same collisional relaxation time
$\tau_{\rm coll}$ for all multipolarities $l \geq 2$:
$\tau_{\rm coll}=\kappa/T^2$,
where $\kappa = 3/(\kappa_2^{-1}+\kappa_3^{-1}+\kappa_\infty^{-1})=628$
MeV$^2$fm/c.
In the limit of large relaxation times the interrelationship
between various sources of dissipation can be ignored \cite{KPS96}
and one can treat also the wall dissipation as an additional source
in the collision term by defining the total relaxation time as
\begin{equation}
   \tau^{-1} = \tau_{\rm coll}^{-1} + \tau_{\rm wall}^{-1}~.  \label{tau}
\end{equation}
Here we neglected the particle emission from an excited nucleus.

Applying the formalism of the linearized Landau-Vlasov equation
in the relaxation time approximation \cite{KPS96,LPCD00}
leads to the following approximate expression for the
imaginary part $\omega_I$ of the giant multipole resonance
frequency $\omega = \omega_R + i \omega_I$~:
\begin{equation}
   \omega_I \simeq -q \omega_R \,
                   \frac{ \omega_R \tau}{ 1 + q (\omega_R\tau)^2 }
\label{omega_I}
\quad ,
\end{equation}
where the factor $q$ is related with the Landau parameter $F_0$ as
$q = 2/5(1+F_0)$.
Eq. (\ref{omega_I}) is, in fact, a suitable interpolation between the two well known
limits \cite{AK59} of rare collisions $\omega_R\tau \gg 1$ (zero sound) with
$\omega_I \simeq -1/\tau$ and of frequent collisions $\omega_R\tau \ll 1$ (first sound)
with $\omega_I \simeq -q\omega_R^2\tau$.

The Landau parameter $F_0$ can be expressed via the nuclear matter incompressibility $K_\infty$
and Fermi energy $E_F=p_F^2/2m_L^*$ as $K_\infty = 6 E_F (1+F_0)$.
The Landau effective mass $m_L^*$ is connected to the Landau parameter $F_1$
as $m_L^* = m \left(1+\frac{1}{3}F_1\right)$.
Using the NL3$^*$ parameter set of the RMF model we obtain
$F_0=-0.20$ and $F_1=-1.04$.

We have applied Eq. (\ref{omega_I}) with $\omega_R=E^\star$ to compute the GMR width
$\Gamma=-2\omega_I$. The result is shown by the solid line in the lower panel
of Fig.~\ref{fig7}. We observe that Eq. (\ref{omega_I}) gives a reasonable
estimate of the magnitude of the collisional broadening for heavy nuclei.
However the GMR width computed by using Eq. (\ref{omega_I}) is too large for medium
and light nuclei. This is expected, given the fact that we have used the upper limit
of the ``apparent temperature'', which becomes unphysically high
for small mass numbers.

We have to point out that the zero sound damping conditions
are valid for medium-to-heavy mass region $A > 50$, where we have $\omega_R\tau > 6$.
This creates a puzzle, since, as we have seen above, the nuclear matter zero sound 
model would strongly overestimate the experimental GMR centroid energies. 
The answer could be, that the finite size effects essentially modify the zero 
sound mode in a real nucleus.


\begin{figure}[t]
\begin{center}
\includegraphics[clip=true,width=0.7\columnwidth,angle=0.]
{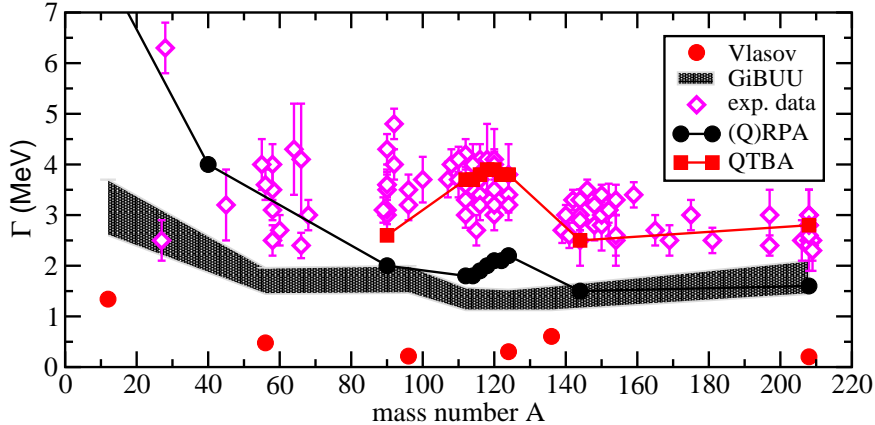}
\caption{\label{fig8} 
(Color online) Width of the GMR as a function of the mass number.
Vlasov and full GiBUU results (filled circles and gray band, respectively)
are compared with experimental data (open diamonds) taken from
\protect\cite{Youngblood}. The solid-circle and
dashed-square curves shows the (Q)RPA and QTBA result, respectively, taken
from Ref. \protect\cite{QTBA}.
}
\end{center}
\end{figure}

The GMR damping mechanism has been a long-winded problem in quantal structure
calculations in the spirit of the RPA \cite{RingSchuck}
and GCM calculations \cite{Sharma09}. It is not the
scope of this work to list all the various structure calculations and discuss
their details, however, they can serve for a qualitative comparison with our
calculations. For more details we refer
to review articles \cite{chomaz,specht}. Fig. \ref{fig8} displays again the
GMR width in the Vlasov approach and the full BUU transport model and
shows a comparison with the two sets of calculations from \cite{QTBA}.

The (Q)RPA calculations, except for the contribution from pairing effects
in open-shell nuclei, can be considered as the quantum analogue of
our semiclassical Vlasov-mode calculations. Collective response is
a superposition of 1p-1h excitations in the both types of calculations.
The difference between (Q)RPA and Vlasov results for the GMR width
stems from missing the quantum fragmentation (Landau damping)
width contribution in our calculations\footnote{In the analytical models
based on BUU equation \cite{KPS96,DKL99} the quantum fragmentation width
contribution has been included as an additional source term in kinetic
equation. However, this is a pure phenomenological way to describe
the damping width of giant multipole resonances in the ground state nuclei.}.
Including a quasiparticle-phonon coupling in the quasiparticle time blocking
approximation (QTBA) \cite{QTBA} can be regarded as a coupling to the
2p-2h configurations \cite{Tselyaev07}, which leads to the strong increase
of the total GMR width. We observe a similar effect in our full BUU calculations,
since nucleon-nucleon collisions in the Uehling-Uhlenbeck collision term
also generate 2p-2h excited states. Moreover, the difference
between QTBA and (Q)RPA results is quite close to the difference between
full BUU and Vlasov results. This again indicates the importance of the correct
description of the fragmentation width contribution.

In the spirit of refs. \cite{Ayik92,Belkacem95}, where the authors argue, that
the Markovian approximation (i.e. a standard Uehling-Uhlenbeck collision term)
is not able to produce any broadening of the giant multipole vibrations at zero
temperature due to severe restrictions of the available phase space for two-body
collisions, our result on the width enhancement by two-body collisions
looks quite surprising. In \cite{Belkacem95}, the authors explain about 25-30\% of
the observed GMR width by taking into account the memory effects in collision term
(non-Markovian approach). However, the analytical models
\cite{KPS96,DKL99,Ayik92,Belkacem95} are based on the linearized kinetic equation
violating the energy conservation. In particular, the temperature increase due
to the damping of collective motion is completely neglected in these approaches.
We stress, that the solution of the full non-linear BUU equation in finite system
leads to much stronger collisional broadening, than expected from the linearized models.

\section{\label{sec5}Conclusions and outlook}

The aim of the present work has been to study theoretically the
excitation energy and the width of the giant monopole resonance
in the framework of a semiclassical transport model.
For this purpose, a description of nuclear ground
states and their phase-space evolution within a unified semiclassical
framework is indispensable. Nuclear ground states are
described by the RTF method using a relativistic Hamiltonian density functional
which, in particular, contains the space-derivatives of the mean
meson fields. The proton and neutron densities from the RTF calculations
serve to generate initial test particle configurations of different nuclei.
The time evolution of the nuclear system is calculated by solving the
kinetic equations which employ the mean-field potentials consistent with those
used in RTF. The improved transport model gives a perfect stability of different
ground state nuclei over long time scales in pure Vlasov dynamics.

The improved initialization method was found to be important to generate
a clear signal of the breathing mode.
Except for light nuclei, pure Vlasov calculations predict a mass dependence
of the excitation energy which is consistent with available experimental
data and with a simple liquid-drop model.
The situation is, however, more complex concerning the lifetime of the
breathing mode.
Vlasov simulations strongly underpredict the experimental data on the GMR
spreading width. The GMR width calculated by using the Vlasov equation behaves
as $\varpropto A^{-1/3}$, which is consistent with the (modified) wall formula.

In order to better understand the GMR damping mechanism, the full transport
calculations of the GMR mode have been performed for the first time.
The inclusion of two-body collisions strongly enhances the
total GMR width to a better agreement with experimental data.
The strong damping of the GMR including the collision term can be understood
in terms of analytical models for one-body and two-body dissipation taking
into account the temperature increase due to the dissipation of the GMR motion.

The Pauli blocking strongly influences the dynamics
in full BUU calculations. Thus it has been treated as precise as possible in
the present work.
The spurious particle emission due to incomplete (numerical) Pauli blocking
destroys the stability of nuclei on the long time scales of the order of ten
periods of GMR oscillations and has been subtracted. This, however,
produce a systematic error of about 30\% in our results on the GMR
damping width.

Overall, the full BUU calculation underestimates the total GMR width by
about $30-50 \%$.
This might be related to the missed Landau damping
contribution in our semiclassical approach, as
the comparison between Vlasov and RPA calculations may indicate.
Moreover, the memory effects in a collision integral neglected in our
calculations also increase the widths of giant multipole resonances.

In conclusion, the breathing mode energies and widths in medium-to-heavy
nuclei are reasonably well described within the GiBUU approach,
when the same Hamiltonian energy functional is consistently used
in the initialization procedure of nuclear ground states and their
phase-space evolution. Thus, an extension of the present work
to investigate other modes of collective excitation,
such as isovector dipole and isoscalar quadrupole resonances,
seems possible and would be a helpful tool to understand better the
dynamics in low energy reaction physics. 
Future transport applications to fusion/deep-inelastic collisions 
for isospin asymmetric systems to investigate exotic collective 
modes in neutron-rich finite systems, such as the pigmy dipole 
resonance, seem possible. 
Furthermore, the improved transport model may be a better tool in
theoretically describing hadron-induced reactions, in-particular the
proton-induced those leading to nuclear fragmentation, and very
peripheral heavy-ion collisions, in which the stability of nuclear
ground states is essential.


{\it Acknowledgments:}
We are grateful to I.N. Mishustin, N. Tsoneva for stimulating
discussions and their interest in this work, and to M. Kaskulov for
his useful suggestions in extracting the width. This work is supported 
by the Helmholtz International Center for FAIR within the framework 
of the LOEWE program and by BMBF.


\end{document}